\newcommand{\keV}{\mathrm{keV}}

\newcommand{\um}{\mu\mathrm{m}}
\newcommand{\nm}{\mathrm{nm}}

\newcommand{\dd}{\mathrm{d}}
\newcommand{\bdd}{\!\!\!\!\mathrm{d}}

\newcommand{\smalltimes}{\scriptsize{\times}}

\newcommand{\cpp}{C\nolinebreak\hspace{-.05em}\raisebox{.4ex}\,\nolinebreak\hspace{-.10em}\raisebox{.25ex}{\footnotesize\bf ++}\,}
\def\cpp{{C\nolinebreak[4]\hspace{-.05em}\raisebox{.25ex}{\footnotesize\bf ++}}\,}
\def\cpp{{C\nolinebreak[4]\hspace{-.05em}\raisebox{.4ex}{\tiny\bf ++}}}

\newcommand{\gff}{g_{F_4}}

\newcommand{\mean}[2]{\big<#1\big>_{#2}}

\newcommand{\loss}{\mathcal{L}}

\newcommand{\xx}{\mathbf{x}}
\newcommand{\rr}{\mathbf{r}}

\documentclass[10pt, a4paper, unknownkeysallowed]{extarticle}
\usepackage[left=2cm,right=2cm,top=2cm,bottom=2cm,includeheadfoot]{geometry}

\usepackage{amssymb}			
\usepackage{amsmath}
\usepackage[makeroom]{cancel}

\usepackage{xcolor}
\usepackage{soul}
\definecolor{green}{RGB}{50, 205, 50}
\definecolor{ptbdunkelblau}{RGB}{0,115,170}
\definecolor{ptbhellblau}{RGB}{0,155,206}
\definecolor{jungmetrologen}{RGB}{240,240,240}

\usepackage{mathtools}
\usepackage{multicol}
\usepackage{multirow}
\usepackage{float}
\usepackage{hyperref}
\usepackage{cite}
\usepackage{comment}
\usepackage{titlesec}

\usepackage[most]{tcolorbox}
\usepackage{graphicx}
\usepackage[export]{adjustbox}

\usepackage[font=normalsize,labelfont=bf, figurename=Fig.]{caption}
\usepackage[subrefformat=parens,labelformat=parens]{subfig}

\DeclareCaptionLabelFormat{mainfigure}{\figurename~#2}
\DeclareCaptionLabelFormat{maintable}{\tablename~#2}
\DeclareCaptionLabelFormat{supplementaryfigure}{Supplementary~\figurename~#2}
\DeclareCaptionLabelFormat{supplementarytable}{Supplementary~\tablename~#2}
\usepackage{natbib}

\let\cite\citep
\let\citeintext\oldcite

\begin{document}
\begin{center}	
	\begin{minipage}{13cm}
		\begin{center}	
			\setlength\parindent{5pt}
			\newcommand{\thetitle}{Active learning-based variance reduction for Monte Carlo
				simulations: A feasibility study for the nanodosimetry around a gold
				nanoparticle}
			\vspace{.5cm}
			\textbf{\Large{\thetitle}}\\
			\vspace{.5cm}
			\small Leo Thomas$^\ast$,
			\small Miriam Schwarze,
			\small Hans Rabus
			\small \\
			\textit{Physikalisch-Technische Bundesanstalt (PTB),
				Abbestr. 2-12,
				D-10587 Berlin, Germany} \\
			\vspace{2mm}
			\hrule
			\vspace{2mm}
			$^\ast$ E-mail: \href{mailto:leo.thomas@ptb.de}{leo.thomas@ptb.de}
			\vspace{.5cm}
		\end{center}
	\end{minipage}
\end{center}

\setcounter{section}{0}

\begin{center}
	\centering
	\begin{minipage}{13cm}

\begin{center}
	\textbf{Abstract}
\end{center}

\textbf{Objective:}
This work presents a data-driven importance sampling-based variance reduction (VR)
scheme inspired by active learning. The method is applied to the estimation of an
optimal impact-parameter distribution in the calculation of ionization clusters
around a gold nanoparticle (NP). Here, such an optimal importance distribution can
not be inferred from principle.
\textbf{Approach:}
An iterative optimization procedure is set up that uses a Gaussian Process Sampler
to propose optimal sampling distributions based on a loss function. The loss is
constructed based on appropriate heuristics. The optimization code obtains estimates
of the number of ionization clusters in shells around the NP by interfacing with a
Geant4 simulation via a dedicated Transmission Control Protocol (TCP) interface.
\textbf{Main results:}
It is shown that the so-derived impact-parameter distribution easily outperforms the
actual, uniform irradiation case. The results resemble those obtained with other VR
schemes but do still slightly overestimate background contributions.
\textbf{Significance:}
While the method presented is a proof-of-principle, it provides a novel method of
estimating importance distributions in ill-posed scenarios. The presented TCP
interface described here is a simple and efficient method to expose compiled Geant4
code to other scripts, written for example, in Python.

	\end{minipage}
\end{center}

\vspace{0.5cm}



\section{Introduction}
\label{section:introduction}


Monte Carlo (MC) simulations are known for their accuracy and are sometimes the only
option to estimate physical quantities. Nanodosimetric calculations, for example,
can---as of now---not be computed analytically.

Such simulations, however, may be computationally expensive so that variance
reduction techniques are necessary. One such approach is importance sampling.
Importance sampling is a technique that improves sampling efficiency by drawing
samples from a so-called \textit{importance distribution} rather than the original
distribution. This works as the bias introduced by sampling from the ``wrong''
distribution can be accounted for easily. The importance distribution can be chosen
freely (as long as it fulfills a criterion discussed later) and ought to have
favorable attributes. A principled heuristic is to choose the importance
distribution so that regions in the input space that contribute more significantly
to the quantity of interest are sampled more frequently.

In MC simulations, however, it is difficult to come up with an appropriate
importance function: the result obtained from MC simulations is not deterministic,
but rather an estimate of an expectation value, i.e. a random variable itself. For
large numbers of contributions to that estimate, the law of large numbers assures
the estimates\textquotesingle \ reliability. When studying a quantity of interest
for sub-regions of the input space---in order to inform an efficient importance
distribution---the estimates may become quite untrustworthy.

In some cases, a second issue occurs: when a quantity of interest consist of a set
of values rather than a single value. The relationship between the input space and
the target quantities may be complicated and difficult to judge. An example is the
calculation of absorbed dose in multiple volumes: a sample drawn from the input
distribution might have different impacts on the scoring volumes. And it is not
immediately clear how the contributions to individual output quantities should be
weighted, effectively rendering the problem under-determined.

This work\textquotesingle s contribution is the introduction of a heuristic that
determines an importance function in an active, data-driven manner, that is: what
contributions are taken into account does not need to be determined a priori but is
inferred from the data obtained. Beyond that the importance function is optimized
for using a loss function as is often used in machine learning. This allows for
straightforward implementation of other optimization objectives that cannot be
implemented easily with other variance reduction methods.


A suitable application is the estimation of the number of ionization clusters in
shells around a gold nanoparticle (NP). The dosimetric properties of gold NPs have
been under review since the studies of \citeintext{Hainfeld_2004} who presented
evidence of their usefulness as radiosensitizers in tumor treatment. To better
understand the effect several works have studied the dose enhancement around a
single uniformly irradiated gold NP \cite{McMahon_2011, Li_2020a, Li_2020b},
demonstrating a locally increased dose enhancement mainly confined to distances of
up to $200\,\nm$ from the NP surface. Using superposition and modelling the spatial
NP distribution in cells, this single NP data can be used to estimate overall cell
survival \cite{McMahon_2011, Lin_2015, Brown_2017, Francis_2019, Velten_Tome_2023,
	Velten_Tome_2024, Rabus_Thomas_2025} using the radiobiological models such as the
local effects model (LEM) \cite{Scholz_1999, Krämer_2000, Elsässer_2007}. More
recently it has been shown that the presence of a gold NP disproportionally enhances
the $F_4$-ionization cluster dose\footnote{There defined as the number of clusters
	consisting of $4$ or more ionizations in a volume divided by the mass of that
	volume.} \citep{Thomas_2024}, emphasizing the necessity for nanodosimetric analysis,
when studying the gold NP enhancement effect.

From the aforementioned studies, only few have exactly addressed the issue of
charged particle equilibrium (CPE): to avoid overestimating the enhancement
effect---often by orders of magnitude---it is essential to properly
account for the background contribution by secondary particles. For the low-kVp
X-rays that are prevalent in gold NP enhancement simulation studies, the secondary
electrons have maximum ranges of several tens of micrometers. Increasing the
diameter of a uniform beam to account for this effect greatly reduces the fluence
incident on the NP in a simulation for the same number of primary particles
simulated.


While the method presented here increases the beam diameter appropriately, it
iteratively uses acquired knowledge to adjust the distribution of the starting
positions of primary particles with the objective of favoring regions according to
their contribution to the ionization cluster dose. For these iteration steps it is
sufficient to simulate $10^6$ primary photons which may be done on a desktop
computer. Once the optimization is concluded, a separate simulation is performed for
more precise results. The CPE issue as well as the method presented here are
detailed in Sections~\ref{subsection:problem} and \ref{subsection:optimization},
respectively.

Within this work the gold NP CPE problem is a use case that serves as a proof of
principle. The method described can be directly used for other spherically symmetric
applications (e.g. diffusing alpha-emitters radiation therapy (DaRT)
\cite{Ballisat_2025}) or adjusted appropriately to other geometries.



A purely practical---though far from irrelevant---aspect of this work is the
interactive implementation of the optimization setup. MC simulations require
computational efficiency in both ``under the hood'' tasks, such as particle
transport, as well as in user code, such as tallies, implementation of geometry and
physics models. Most MC codes are implemented as (precompiled) software libraries in
Fortran or \cpp. Modern statistical applications on the other hand usually also
emphasize quick development time, relying on higher-level languages such as Python,
and rely on precompiled software libraries (such as Numpy, Scipy or Pytorch) only
for intensive tasks\footnote{An exception are recent developments within OpenGATE
	\cite{Sarrut_2014, Sarrut_2021, Sarrut_2022} that aim to expose (compiled) Geant4
	functionalities to Python code. As of now, this library, however, is not yet capable
	of offering the advanced functionalities required for this application, such as
	track structure computation or applying different physics-models to different
	simulation volumes.}.

While it is usually feasible to run a simulation and perform further analysis later
on, a data-driven approach such as the one presented here requires the optimization
component to interactively receive MC-derived estimates (such as the cluster dose in
shells around a NP) for certain input parameters (such as a range of impact
parameters). Here this is done using the Transmission Control Protocol (TCP)
interface that native to all major operating systems, allowing for the simulation
code to stand-by for requests, as opposed to needing to initialize for each
computation, a process that often takes tens of seconds and can lead to significant
overhead for iterative tasks.

\section{Methods}
\label{section:methods}


\subsection{Introduction to the problem}
\label{subsection:problem}

\subsubsection{Charged particle equilibrium}
\label{subsubsection:cpe}

\begin{figure}[t]
	\centering
	\fbox{
		\begin{minipage}{0.667\textwidth}
			\vspace{5 mm}
			\begin{center}
				\includegraphics[width=0.75 \textwidth]{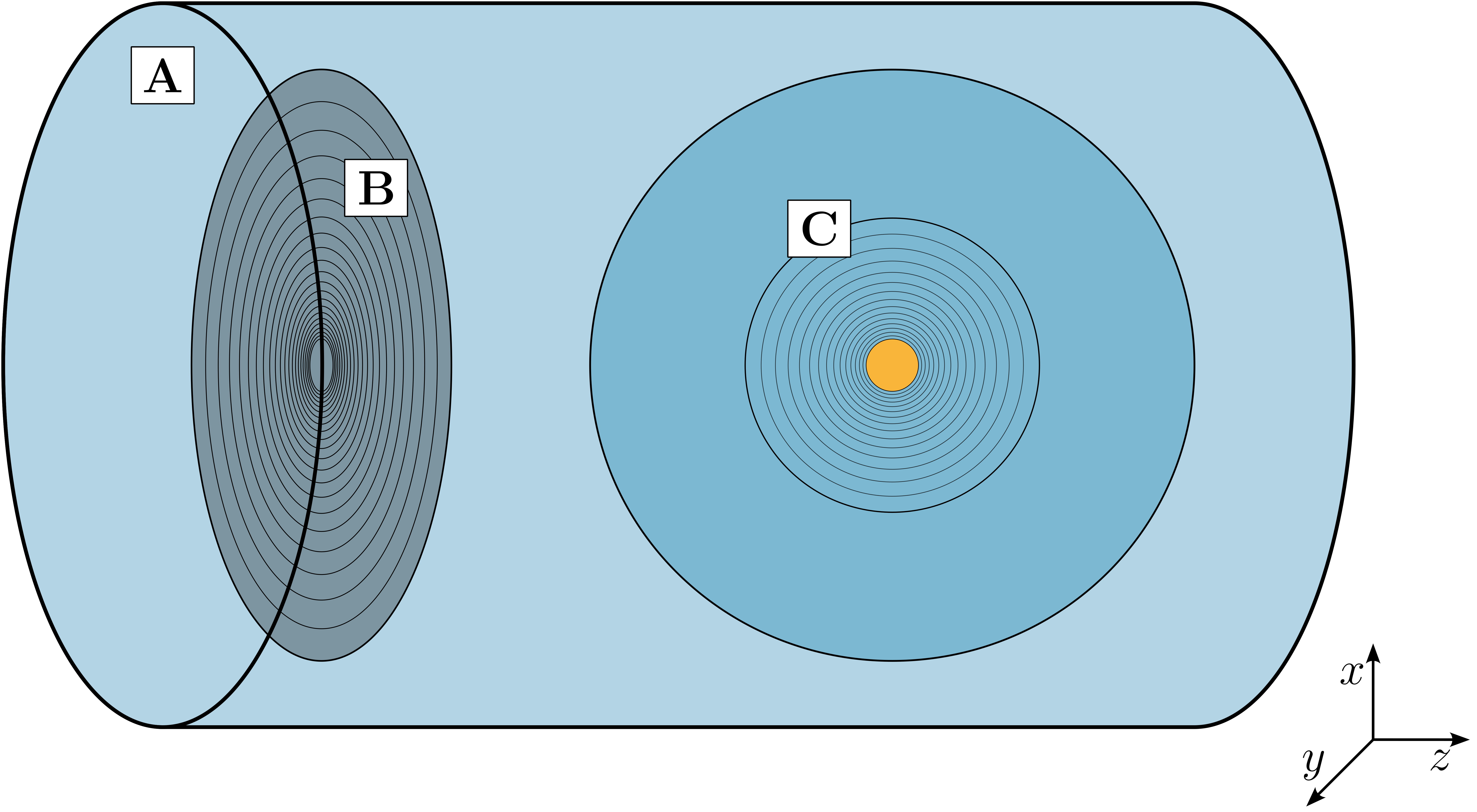}
			\end{center}
			\caption{Geometrical irradiation setup (not to scale, c.f. Fig.~\ref{fig:impact}
				for a more detailed description of the geometry).
				The world volume (region `$\mathbf{A}$') is a
				cylinder filled with water. Located at its center is a gold NP (in
				yellow). The source has the shape of a disk, located in the
				$x$-$y$-plane at a distance of $100\,\um$ from the NP center. It is
				divided into annuli that are logarithmically increasing in width
				(region `$\mathbf{B}$'). The dark blue volume (region `$\mathbf{C}$')
				is the volume in which secondary electrons are transported. Its size is
				chosen so that any electron track that could reach any of the scoring
				shells around the NP is included.
			}
			\label{fig:setup}
		\end{minipage}
	}
\end{figure}

The objective in this work is the calculation of the $F_4$-cluster dose $\gff$
(after \citeintext{Faddegon_2023}), defined as the number of ionization clusters
consisting of four or more ionizations per mass in radial shells around a gold NP
under X-ray irradiation. The calculation is carried out with a simulation of a
single gold NP in water (see Fig.~\ref{fig:setup}). The details of the simulation
are elaborated on in section~\ref{subsection:simulation}.

A simulation of the irradiation of a single gold NP under CPE conditions requires
the generation of primary particles covering an area that allows for contributions
of secondary particles originating from a distance to be considered. For the
$100$-kVp X-ray spectrum considered, the maximum range such electrons can travel is
estimated to be $d_\mathrm{max} = 50\,\um$. This is roughly equal to the Continuous
Slowing Down Approximation (CSDA) range of electrons in water with a kinetic energy
of $54.4\,\keV$ \cite{ICRU_37}. For a radiation field corresponding to the
irradiation setup chosen here it has been shown that $97.2\,\%$ of the produced
secondary electrons have a kinetic energies below that \cite{Thomas_2024}.

When the starting positions of the simulated primary particles are distributed over
such a large area it leads to a drastic reduction of the effective incident fluence
$\phi$, defined as the number of primary particles $N$ incident on a sphere per the
cross-sectional area of that sphere \cite{ICRU_85}. For a `narrow' beam, collimated
to the extension of a NP, the fluence is given by
\begin{align*}
	\phi_{\scriptstyle\mathrm{narrow}} = \frac{N}{r_\mathrm{np}^2\pi}\,.
\end{align*}

If one considers a sphere of radius $r_\mathrm{roi} > r_\mathrm{np}$ as region of
interest (ROI) within which secondary particle equilibrium conditions shall be fulfilled
the fluence will reduce dramatically. For a NP of radius $r_\mathrm{np} = 50\,\nm$
and $r_\mathrm{roi} = 5\,\um$, the fluence is reduced by a factor of
\begin{align*}
	\frac{\phi_{\scriptscriptstyle\mathrm{CPE}}}{\phi_{\scriptstyle\mathrm{narrow}}}
	\ =\ \left(\frac{r_\mathrm{np}}{r_\mathrm{roi} + d_\mathrm{max}}\right)^2
	\ \approx\ 8.3\smalltimes 10^{-7}.
\end{align*}

Clearly, obtaining results with agreeable precision requires some ingenuity. One method
to address the problem of the bias introduced by using a beam
confined to the NP dimensions and correcting the resulting lack of secondary particle
equilibrium by estimating the collision kerma in water \cite{Rabus_2019,
	Rabus_Li_2021}. A more common strategy is to split the simulation into two (or more)
steps. This allows to analyze the radiation field that would be incident on the NP
surface (or in its vicinity) and estimate radiation effects in a later step. This
method allows for the application of variance reduction schemes or other
simplifications. Such methods often use phase-space files \cite{Lin_2014,
	Velten_Tome_2023, Klapproth_2021, Taheri_2025} or estimate the spectral fluence of
particles present at the NP \cite{Thomas_2024}. While generally sound, the two-step
method usually relies on simplifications, such as ignoring synergistic effects of
secondary particles originating from the same primary particle.


\begin{figure}[t]
	\centering
	\fbox{
		\begin{minipage}{0.667\textwidth}
			\vspace{5 mm}
			\begin{center}
				\includegraphics[width=0.75 \textwidth]{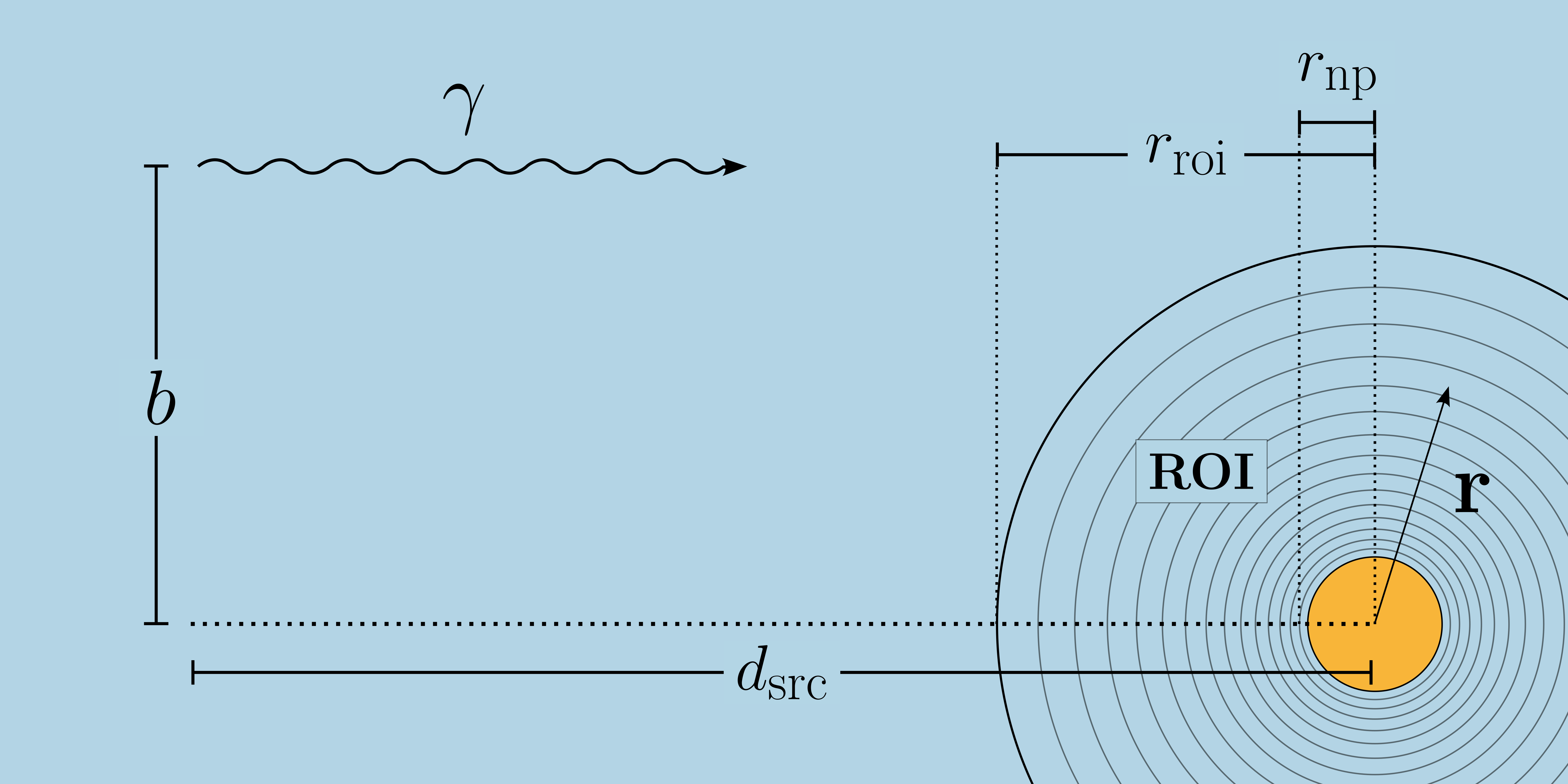}
			\end{center}
			\caption{Illustration of the coordinate system used with a gold NP (in
				yellow) placed at the origin. The shells around the NP symbolize the scoring
				volumes. Collectively they make up the region of interest.
				The source is a disk located in the $x$-$y$-plane at
				$z=-d_\mathrm{src} = -100\,\um$ and photons are generated with a
				momentum direction of $(0,0,1)$.
			}
			\label{fig:impact}
		\end{minipage}
	}
\end{figure}

\subsubsection{Expectation value perspective}

The method presented here takes a different way of approach, and its development
requires some formalization. MC simulation calculates physical observations as
ensemble averages over multiple particle trajectories and is, at its core, the
estimation of an expectation value of an observable. The quantities influence that
is of interest in the present case is the primary particle position. In fact, due to
the cylindrical symmetry of the setup it is only the lateral displacement from the
central axis, the impact parameter $b$, that is considered (see
Fig.~\ref{fig:impact}).

The observable of interest here is the $F_4$ ionization cluster dose $\gff$ as a
function of radial distance from the NP normalized to the primary photon fluence
$\phi_0$. It is the expectation of an---not directly accessible---function
$\gff(r|b)$ with regard to the impact parameter distribution $p(b)$:
\begin{align}
	\frac{\gff(r)}{\phi_0}
	\ =\ \frac{1}{\phi_0}\underset{b \sim p}{\mathbb{E}}
	\Big[\gff(r|b)\Big]\ =\ \frac{1}{\phi_0} \int\dd b\ p(b)\gff(r|b), \label{eq:exp-continuous}
\end{align}
where $\gff(r|b)$ is the $F_4$-cluster dose at a distance $r$ from the NP center,
given that the primary particle originates from a point with impact parameter $x$.

The conditional dependence emphasizes the distinction between the two variables:
while $r$ is the variable whose influence on $\gff$ is to be studied, the impact
parameter $b$ serves as an input parameter needed to compute that relationship. Here
$x$ is modelled as a stochastic variable, while $r$ is a deterministic variable.
Effectively it is this expectation value in eq.~\ref{eq:exp-continuous} that is
estimated via simulation:
\begin{align*}
	\int\dd b\ p(b)\gff(r|b)
	\ \approx\ \frac1N\sum_{k=1}^N \gff(r|b_k)\quad\mathrm{with}\ b_k \propto p\
	\forall\ k=1,\dots , N,
\end{align*}
where $N$ is a number of samples drawn from $p$.

The expectation in eq.~\ref{eq:exp-continuous} is an idealization in form of a
continuous function; in
practice, the ionization cluster doses are scored as averages over spherical shells.
Similarly, the impact parameter space is divided into intervals (which form annuli
on the disc-shaped source). Within an annulus, primary position are then sampled
uniformly. This is known as \textit{stratified sampling} and effectively allows
treating the distribution of impact parameters as discrete, facilitating later
optimization. The corresponding contributions to the cluster dose are denoted by a
mean $\mean{\gff}{r_i, b_j}$, which is the average cluster dose $\gff$ in a shell
defined by $r\in [r_i, r_{i+1})$ originating from primary particles with impact
parameter $x\in [b_j, b_{j+1})$\footnote{Note: For comprehensibility in this work
	the index $i$ will always refer to a radial shell whereas the index $j$ will always
	refer to an impact parameter annulus.}. Eq.~\ref{eq:exp-continuous} becomes then:
\begin{align}
	\frac{\mean{\gff}{r_i}}{\phi_0}
	\ =\ \frac{1}{\phi_0}\sum_{j=0}^{n_b-1}p_j\,\mean{\gff}{r_i, b_j}
	\label{eq:exp-mean}
\end{align}
where $n_b$ is the number of shells and $p_j$ refers to the \textit{probability
	mass} in an annulus (defined in eq.~\ref{eq-a:masses}). The derivation of eq.~\ref{eq:exp-mean} is detailed in
appendix~\ref{subsection:averaging}.

When gathering samples to estimate the cluster dose around the NP, many of the
primary particles contributions to scoring may be insignificant but do make the
computation effort immense. This poses the questions:
\begin{itemize}
	\item To what extent do contributions from a given impact parameter affect the quantity of
	      interest?
	\item And can this information be practically used for variance reduction?
\end{itemize}

\subsubsection{Importance sampling}
\label{subsubsection:importance}

Importance sampling is a suitable technique here: expectations such as the one in
eq.~\ref{eq:exp-continuous} can be evaluated by drawing samples from a probability
distribution other than the actual physical distribution arising from the setup: the
so-called \textit{importance distribution}. Favoring impact parameters while
discouraging others leads to a bias of the sampling process. This bias, however, can
be accounted for:
\begin{align*}
	\frac{\gff(r)}{\phi_0}
	\ =\ \frac{1}{\phi_0}\underset{b \sim p}{\mathbb{E}} \Big[\gff(r|b)\Big]
	\ =\ \frac{1}{\phi_0}\underset{b \sim q}{\mathbb{E}}
	\left[\frac{p(b)}{q(b)}\gff(r|b)\right].
\end{align*}
The term $p(b)/q(b)$ is referred to as \textit{likelihood ratio}. The importance
distribution $q(b)$ can be chosen arbitrarily as long as
$\mathrm{supp}(q) \supseteq \mathrm{supp}(p)$. Importance sampling can be applied
to the annulus- and shell-averaged quantities that are estimated here (see
appendix~\ref{subsection:importance}), so that eq.~\ref{eq:exp-mean} becomes
\begin{align}
	\frac{\mean{\gff}{r_i}}{\phi_0}
	\ =\ \frac{1}{\phi_0}\sum_{j=0}^{n_b-1} q_j \frac{p_j}{q_j}\,\mean{\gff}{r_i,
		b_j}, \label{eq:exp-mean-imp}
\end{align}
where $q_j$ is the probability mass associated with the importance function $q(b)$,
defined in eq.~\ref{eq-a:masses-q}. In fact eq.~\ref{eq:exp-mean-imp} is identical
to eq.~\ref{eq:exp-mean}, as the $q_j$ simply cancel out. The reason for this is
that $\mean{\gff}{r_i, b_j}$ is already independent of the number of primary
particles and in principle converges to the same mean for any distribution---though
the rates of this convergence may vary considerably. In the case of an expectation
value of a single-valued variable, it is possible to select an optimal candidate for
the importance function $q$ where `optimal' refers to the choice for the importance
function that minimizes the variance of the estimator obtained by a Monte Carlo
sampling approach. It can be shown that the optimal choice for the importance
function $q(b)\propto p(b)|f(b)|$ if $f(b)$ is the random variable for which the
expectation-value is to be estimated (cf. for example theorem 3.12 in
\citeintext{RobertCasella_2004}). The heuristic is simple: increase sampling where
the contributions are the largest.

The referenced theorem, however, does not straight-forwardly lend itself to the case
at hand: it applies to the expectation of a deterministic function, from which
accurate values can be obtained. Here, $\gff$ is a stochastic quantity itself that
can only be estimated.

A second issue is that the observable $\left<\gff\right>_{r_i}$ is a multi-variate
random variable: it is unlikely that there is a distribution $q$ that will minimize
the variance of the estimator of the cluster dose in \textit{all} shells. This case
corresponds to a multi-objective optimization and in that sense, the problem is
over-determined and the heuristic needs to be adjusted appropriately.

\subsection{Active Optimization of Importance Sampling}
\label{subsection:optimization}

Starting point for such an adjusted heuristic is the contribution that an annulus
$j$ has to a shell $i$. While this relation is a priori non-trivial, the information
can appropriately be encapsulated in a matrix. The following quantity is proposed:
\begin{align*}
	W_{ij} \equiv \frac{p_j\mean{\gff}{r_i, b_j}}{\sum_{k=0}^{n_b-1} p_k\mean{\gff}{r_i,b_k}}
	\ \overset{(\ref{eq:exp-mean})}=\ \frac{p_j\mean{\gff}{r_i,
			b_j}}{\mean{\gff}{r_i}}.
\end{align*}

$W$ will be referred to as the \textit{contribution matrix}. It is the relative
contribution of primary particles generated on annulus $j$ (i.e. the annulus
delimited by $[b_j, b_{j+1})$) to the cluster dose in shell $i$. The denominator is
chosen so that $\sum_{j=0}^{n_b-1} W_{ij} = 1\ \forall\ i=0,\dots,n_r-1$, i.e. the
sum of the relative contributions of all annuli to the cluster dose in shell $i$ is
one.
The contribution matrix can be used to project the shell-mean of the cluster dose
$\mean{\gff}{r_i}$ onto single annuli. The so-constructed quantity will be referred to
as the \textit{importance score}:
\begin{align}
	u_j \equiv \sum_{i=0}^{n_r-1} W_{ij} \mean{\gff}{r_i}. \label{eq:u_j}
\end{align}

This importance score is a measure for how relevant the contribution of annulus $j$
is across all shells. This naturally leads to a possible choice for importance
function:
\begin{align}
	q^u_j \propto u_j. \label{eq:qu}
\end{align}

This is in the spirit of the heuristic described in
section~\ref{subsubsection:importance}: Annuli with higher importance scores, that
is, higher contribution to the cluster dose are favored. In practice, values of
$u_j$ are only as informative as the samples of $\mean{\gff}{}$ obtained from
simulation. Especially in earlier iterations, these estimates can be quite imprecise
and thus divert the algorithm. This is remedied by convolving the importance scores
\begin{align}
	u_j \rightarrow (G \ast u)_j, \label{eq:conv}
\end{align}
where $G$ is a discrete Gaussian kernel\footnote{Note: Mathematically,
	eq.~\ref{eq:conv} is not a convolution as the kernel values refer
	to equidistant bins but are applied to data with uneven bin-spacing, see
	eq.~\ref{eq:b_lower} in section~\ref{subsection:simulation}. For
	the sake of optimization, however, different probability masses
	$q_j$ are simply a set of values and applying eq.~\ref{eq:conv}
	enforces smoothness between importance scores nonetheless.}. This
smoothens out the weights, in agreement with the intuition that the importance
score between neighboring annuli should be similar in magnitude.

\subsubsection{Loss function}

In principle, it is possible to compute importance scores from the estimates of
$\mean{\gff}{r_i, b_j}$ via a preliminary simulation run for which an initial
importance function is chosen as $q_j = p_j\ \forall\ j$ and then choose an updated
importance function according to eq.~\ref{eq:qu}.

These importance scores, however, are based on estimators of $\mean{\gff}{r_i,
		b_j}$. In the case at hand, generating reliable estimators is computationally
expensive. In fact, without the use of importance sampling (as is done in a
preliminary run where $q_j = p_j$), such estimators are not precise enough to carry
much information. Therefore, here the condition eq.~\ref{eq:qu} is enforced more
indirectly by using a \textit{loss function}. A loss function $\mathcal{L}$ is a
quantity that reflects the difference between a target set of parameters and a set
of proposed parameters and is used as an objective for minimization. The optimal
importance function is then
\begin{align*}
	\arg\min_{q^\prime\in\mathcal{Q}}\loss(q^\prime),
\end{align*}
with $\mathcal{Q}=\{q\, |\, \mathrm{supp}(q) \supseteq \mathrm{supp}(p)\}$.

While a simple loss function to enforce the condition eq.~\ref{eq:qu} could be
constructed using an L2-norm or a Kullback-Leibler divergence, the Wasserstein-1
distance $W_1$ is used. Unlike the Kullback-Leibler divergence, the Wasserstein
distance is a metric for probability distributions in the mathematical sense and is
deemed a more robust measure: the loss function ought to quantify the difference
between two probability distributions, $q^u$, the importance function implied by the
importance scores (eq.~\ref{eq:qu}) and some candidate importance function
$q^\prime\in\mathcal{Q}$. A loss function can be constructed as
\begin{align}
	\loss_{W_1}(q^\prime) & = \frac{W_1(q^u, q^\prime)}{b_\mathrm{max}}. \label{eq:loss-W}
\end{align}

$W_1$ as a distance between distributions
over the impact parameter space and hence scale dependent. It is normalized to the
maximum impact parameter $b_\mathrm{max}$ to obtain a dimensionless quantity.
The calculation of the $W_1$-distance used is elaborated in 
appendix~\ref{sec:appendixB}.

While evidently an optimal solution for the loss function in eq.~\ref{eq:loss-W} is
$q_j^\prime = q^u_j$, the loss optimization approach, however, offers a greater
flexibility than setting $q_j = q^u_j$: It naturally accommodates the addition of
further terms to enforce additional constraints such as regularization. Balancing
these terms allows to enforce single optimization objectives more or less
aggressively.

Reflecting prior knowledge of the homogeneous geometry in large domains of the
impact parameter space, it makes sense to add a regularization term to the loss
function. This is done by penalizing differences between neighboring distribution
values:
\begin{align*}
	\mathcal{L}_\mathrm{reg}(q^\prime) = \frac{\sum_{j=1}^{n_b-2} (q_{j+1}^\prime -
    q_j^\prime)^2}{\sum_{j=1}^{n_b-2} {q_j^\prime}^2},
\end{align*}
whereby the denominator serves as a normalization. Effectively, this term is
equivalent to minimizing the first derivative of $q$ w.r.t. $b$. Note that the
distribution value corresponding to the first annulus, which matches the NP
dimension is omitted here, as the gold NP renders the geometry inhomogeneous and
smooth behavior is thus not expected. The resulting loss function is
\begin{align}
	\loss(q^\prime) & =
	\loss_{W_1}(q^\prime)
	+ \lambda \mathcal{L}_\mathrm{reg}(q^\prime). \label{eq:loss-final}
\end{align}

Here, $\lambda$ is a parameter that sets the relative magnitude of the
regularization term and the $W_1$-term is normalized to the maximum impact parameter.
Commonly numerical values for hyperparameters such as $\lambda$ are
determined through a separate optimization. Doing so requires a (quantitative)
performance metric to evaluate hyperparameter which is not straight-forward to define
here (this is discussed in section~\ref{section:discussion}). In the present study
$\lambda = 8\smalltimes 10^{-2}$ has been found to limit the
regularization loss\textquotesingle s contribution to the total loss to $\lesssim
	30\,\%$.

The minimization of the loss is done using the Gaussian Process (GP) Sampler by
Optuna \cite{optuna_2019}, an algorithm that is well-suited for the optimization of
probability distributions. It performs a GP regression on the loss function in
parameter space and uses an acquisition function to suggest sets of probability
masses $q_j$ that minimize the loss function.


\subsubsection{Workflow}

\begin{figure}[t!]
	\centering
	\fbox{
		\begin{minipage}{0.667\textwidth}
			\vspace{5 mm}
			\begin{center}
				\includegraphics[width=0.75 \textwidth]{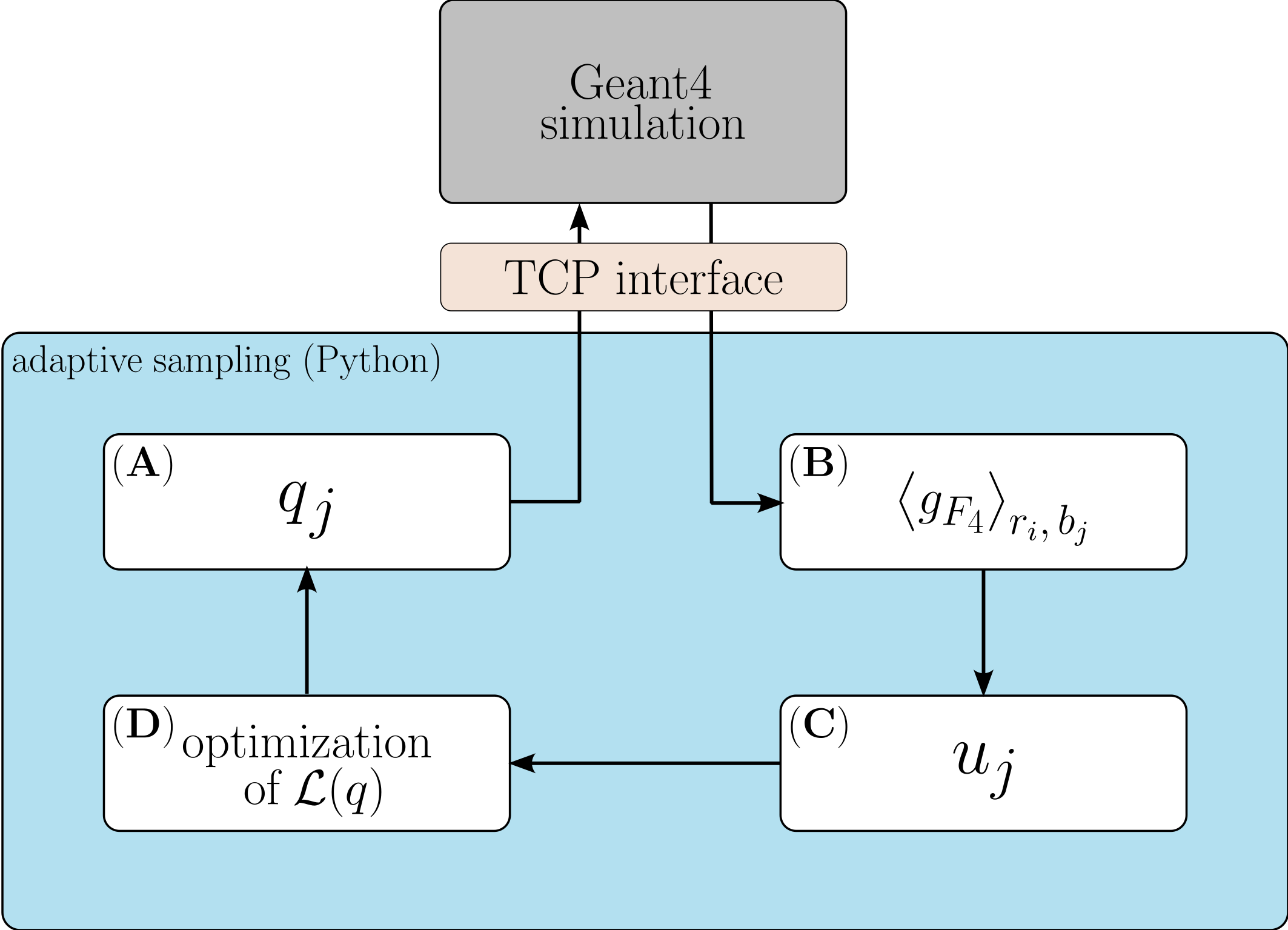}
			\end{center}
			\caption{Flowchart of the optimization procedure. The process begins with
				an initial choice for the importance function $q$ ($\mathbf{A}$). For
				each annulus, a number of primary particles $N^q_j$ is simulated to obtain
				estimates for the annulus- and shell-mean of the cluster dose
				$\mean{\gff}{r_i,b_j}$ ($\mathbf{B}$). Using this information, importance
				scores and the loss function are calculated using eqs.~\ref{eq:u_j} and
				\ref{eq:loss-final}, respectively ($\mathbf{C}$). Using the TPE sampler a
				new importance function $q$ is chosen that optimizes this loss function
				($\mathbf{D}$).
			}
			\label{fig:flowchart}
		\end{minipage}
	}
\end{figure}


Starting point of the optimization is an initial guess of the weights $\{q_j\}_j$
(with $j=0,\dots,n_b-1$), see $\mathbf{(A)}$ in fig.~\ref{fig:flowchart}. This
initial guess will be the physical probability weights $\{p_j\}_j$. Using these
masses, a corresponding number of primary particles
\begin{align*}
	N^q_j = \lfloor q_j N \rfloor
\end{align*}
is generated in each annulus\footnote{$\lfloor \cdot \rfloor$ represents the
	floor function which always returns the largest integer that is smaller than its
argument.}, where the starting positions are uniformly sampled on
the annulus $j$. $N$ is the total number of primary particles generated in all
annuli. In the present work, it has been chosen as $N = 10^6$.
The simulation is done using Geant4-code and a TCP interface, further
detailed in Section~\ref{subsection:simulation}.

For the given values for $\mean{\gff}{r_i,b_j}$, the importance scores $u_j$ are
calculated according to eq.~\ref{eq:u_j} and transformed according to
eq.~\ref{eq:conv}. These are used to construct the loss function, which is
subsequently optimized. Once a new importance distribution $q_j^\mathrm{opt}$ is
determined using the GP sampler, it is used to update the existing importance
function:
\begin{align}
	q_j^{(k)} = \alpha q_j^\mathrm{opt} + (1 - \alpha) q_j^{(k-1)}\quad\mathrm{with}\
	\alpha\in (0,1]. \label{eq:q_update}
\end{align}

This method of updating the importance distribution increases stability between
iterations and prevents outliers to divert the convergence. The distribution in
eq.~\ref{eq:q_update} is then used to generate new cluster dose data. The iteration
is continued until a stationary distribution is obtained. Once the optimization
routine is finished, the obtained importance distribution is used in a separate
comparison simulation with a higher number ($N = 10^8$) of primary particles for
higher precision.

\subsection{Simulation}
\label{subsection:simulation}

The purpose of the simulation is to calculate the $F_4$-ionization cluster dose
around the gold NP for a given impact parameter range. It is designed to yield
accurate estimates that are made available to the optimization code via a dedicated
TCP interface (Section~\ref{subsubsection:tcp}).

The simulation\textquotesingle s design allows for comparison of the resulting data
with data from the \mbox{EURADOS} intercomparison
\cite{Li_2020a,Li_2020b,Rabus_Li_2021,Rabus_2021}, a multi-center comparison of
Monte Carlo codes.

The geometry consists of a cylindrical world volume (region `$\mathbf{A}$' in
Fig.~\ref{fig:setup}) of length $220\,\um$ and radius $100\,\um$. Located at its
center, a $50\,\nm$-radius gold NP is placed.




The primary particles are photons with a kinetic energy corresponding to a $100$-kVp
X-ray spectrum\footnote{The spectrum used in \citeintext{Li_2020a} has been modified
	by removing a (physically implausible) peak between $85.0\,\keV$ and $85.5\,\keV$ by
	averaging over the two adjacent bins.}. The source lies in the $x$-$y$-plane (at $z
	= -d_\mathrm{src} = - 100\,\um$) and is split into different annuli centered around
the $z$-axis (region `$\mathbf{B}$' in Fig.~\ref{fig:setup}). With the exception of
the first annulus which is matched to the gold NP radius ($50\,\nm$), the radii
increase in size exponentially Beginning from the $50\,\nm$ the annulus-size
increases with $10$ bins per decade, so that the positions of the lower bin edges
are given by
\begin{align}
	b_0^\mathrm{lower} & = 0
	\qquad\mathrm{and}\qquad
	b_j^\mathrm{lower}  = 50\,\nm \cdot 10^{(j-1)/10} \label{eq:b_lower}
\end{align}
with $j = 1,\dots, n_b$ and $n_b = 31$ ($3$ decades with $10$ annuli per decade +
the first annulus). A set of primary particles is generated with a starting position
uniformly distributed on a given annulus.
Table~\ref{tab:p_j} lists the lower annulus edges together with the associated
probability masses $p_j$ that correspond to uniform irradiation. The table
illustrates the small fraction of primary particles incident on the gold NP itself as
well as its immediate surrounding.

\begin{table}[thb]
	{\footnotesize
		\begin{minipage}{\textwidth}
			\captionsetup{justification=raggedright,
				singlelinecheck=false
			}
			\caption{Values of $p_j = A_j / A$ (see eq.~\ref{eq-a:masses} in the appendix)
				for the lower edges of the annuli defined in
				eq.~\ref{eq:b_lower} (Note that $b_j^\mathrm{upper} = b_{j+1}^\mathrm{lower}$).
				For a simulation of $N$ primary photons,
				$\lfloor p_j N\rfloor$ photons are generated on annulus $j$.
				This table illustrates the need for variance reduction: In a
				simulation of $N = 10^6$ primary particles, only $1$ primary photon
				is expected to head for
				the NP directly. Given the low interaction probability of photons of
				$100$ kVp X-rays, such a simulation is unlikely to produce particles
				that interact with the NP at all.
				In fact, given the probability of a photon interacting within a
				$50\,\nm$-gold-NP of roughly $1.5\smalltimes 10^{-3}$
				\cite{Thomas_2024}, a number of $N \approx 2/3 \smalltimes 10^{9}$ primary
				particles would yield one expected interaction within the gold NP.
			}
			\begin{center}
				\def\arraystretch{1.5}
				\begin{tabular}{c|c|c||c|c|c||c|c|c}
					$j$ & $b_j^\mathrm{lower}$ / nm & $p_j$                     & $j$ & $b_j^\mathrm{lower}$ / nm & $p_j$                     & $j$ & $b_j^\mathrm{lower}$ / nm & $p_j$                     \\ \hline\hline
					0   & $0$                       & $10^{-6}$                 &     &                           &                           &     &                           &                           \\ \hline
					1   & $5.00\smalltimes 10^{1}$  & $5.85\smalltimes 10^{-7}$ & 11  & $5.00\smalltimes 10^{2}$  & $5.85\smalltimes 10^{-5}$ & 21  & $5.00\smalltimes 10^{3}$  & $5.85\smalltimes 10^{-3}$ \\ \hline
					2   & $6.29\smalltimes 10^{1}$  & $9.27\smalltimes 10^{-7}$ & 12  & $6.29\smalltimes 10^{2}$  & $9.27\smalltimes 10^{-5}$ & 22  & $6.29\smalltimes 10^{3}$  & $9.27\smalltimes 10^{-3}$ \\ \hline
					3   & $7.92\smalltimes 10^{1}$  & $1.47\smalltimes 10^{-6}$ & 13  & $7.92\smalltimes 10^{2}$  & $1.47\smalltimes 10^{-4}$ & 23  & $7.92\smalltimes 10^{3}$  & $1.47\smalltimes 10^{-2}$ \\ \hline
					4   & $9.98\smalltimes 10^{1}$  & $2.33\smalltimes 10^{-6}$ & 14  & $9.98\smalltimes 10^{2}$  & $2.33\smalltimes 10^{-4}$ & 24  & $9.98\smalltimes 10^{3}$  & $2.33\smalltimes 10^{-2}$ \\ \hline
					5   & $1.26\smalltimes 10^{2}$  & $3.69\smalltimes 10^{-6}$ & 15  & $1.26\smalltimes 10^{3}$  & $3.69\smalltimes 10^{-4}$ & 25  & $1.26\smalltimes 10^{4}$  & $3.69\smalltimes 10^{-2}$ \\ \hline
					6   & $1.58\smalltimes 10^{2}$  & $5.85\smalltimes 10^{-6}$ & 16  & $1.58\smalltimes 10^{3}$  & $5.85\smalltimes 10^{-4}$ & 26  & $1.58\smalltimes 10^{4}$  & $5.85\smalltimes 10^{-2}$ \\ \hline
					7   & $1.99\smalltimes 10^{2}$  & $9.27\smalltimes 10^{-6}$ & 17  & $1.99\smalltimes 10^{3}$  & $9.27\smalltimes 10^{-4}$ & 27  & $1.99\smalltimes 10^{4}$  & $9.27\smalltimes 10^{-2}$ \\ \hline
					8   & $2.51\smalltimes 10^{2}$  & $1.47\smalltimes 10^{-5}$ & 8   & $2.51\smalltimes 10^{3}$  & $1.47\smalltimes 10^{-3}$ & 28  & $2.51\smalltimes 10^{4}$  & $1.47\smalltimes 10^{-1}$ \\ \hline
					9   & $3.15\smalltimes 10^{2}$  & $2.33\smalltimes 10^{-5}$ & 19  & $3.15\smalltimes 10^{3}$  & $2.33\smalltimes 10^{-3}$ & 29  & $3.15\smalltimes 10^{4}$  & $2.33\smalltimes 10^{-1}$ \\ \hline
					10  & $3.97\smalltimes 10^{2}$  & $3.69\smalltimes 10^{-5}$ & 20  & $3.97\smalltimes 10^{3}$  & $3.69\smalltimes 10^{-3}$ & 30  & $3.97\smalltimes 10^{4}$  & $3.69\smalltimes 10^{-1}$ \\ \hline
				\end{tabular}
			\end{center}
			\label{tab:p_j}
		\end{minipage}
	}
\end{table}

Simulations were performed using the Geant4-DNA-library (Version 11.2.2)
\cite{Incerti_2010a, Incerti_2010b, Incerti_2018, Bernal_2015, Sakata_2019}. In
region `$\mathbf{C}$', Option 4 models were used for electron transport below
$10\,\keV$ and Option 2 models above. Secondary electrons were not tracked within
region `$\mathbf{A}$'. The spherical region `$\mathbf{C}$' has a radius of
$55.05\,\um$ that is $5\,\um + 50\,\nm$, the largest radius in which cluster dose is
scored, plus $50\,\um$, the upper bound to the secondary electron\textquotesingle s
range.


\subsubsection{Scoring and ionization clustering}
\label{subsubsection:scoring}

Scored were ionizations, namely those produced by electron impact ionization,
photoelectric absorption and incoherent scattering of photons, and non-radiative
de-excitation of core holes.


After each primary track\textquotesingle s calculation was concluded, the ionization
points were clustered using the Associated Volume Clustering (AVC) approach
\cite{Kellerer_1985, Famulari_2017} as
implemented by \citeintext{Thomas_2024}. AVC can be seen as a variant of uniform
sampling of scoring volumes where sampling volumes (the \textit{associated volumes})
are randomly sampled so that they always contain at least one ionization. Here,
$F_4$ clusters are scored, that is, a cluster consists of at least four ionizations.


The resulting ionization clusters are scored as ionization cluster dose on shells
whose thickness increases logarithmically with $20$ bins per decade, so that:
\begin{align}
	r_i^\mathrm{lower} = 50\,\nm \cdot 10^{i/20} \label{eq:r_lower}
\end{align}
with $i = 0,\dots, n_r - 1$ and $n_r = 40$ ($2$ decades with $20$ shells per decade).

\subsubsection{TCP interface}
\label{subsubsection:tcp}

To allow for interactive control of the simulation from an external optimization
routine, the Geant4 simulation is extended with a TCP-based communication interface
using ZeroMQ \cite{Hintjens_2013, zeromq}. TCP (Transmission Control Protocol) is a
low-level communication protocol that allows for data to be exchanged between
programs. Commonly, TCP is used to establish network connections; it can, however,
be used locally just as well. ZeroMQ is an API (Application Programming Interface)
that provides language bindings for various programming languages, including \cpp\
and Python and relieves the user from directly working with sockets on the operating
system level.

In this work, the simulation program acts as a TCP server. It remains in an infinite
loop, continuously waiting for incoming requests. The client (here the optimization
code), is able to connect to the server and send these requests. Each request
specifies a number of primary particles to be generated, along with the lower and
upper bounds of the impact parameter interval (as defined in eq. \ref{eq:b_lower}).
Upon reception of a request, the server simulates a corresponding number of primary
particles on the specified impact parameter range, calculates the $F_4$-cluster
doses in the spherical shells defined by eq.\ref{eq:r_lower} and sends this data
back to the optimization code.

Sending data back and forth between two programs requires serialization and
deserialization. While data within the program is structured, such as an array of
doubles or class-instances, TCP passes data byte by byte. The transformation of
structured data into a linear byte stream and its subsequent reconstruction is
referred to as serialization and deserialization, respectively. While it is
certainly possible to interpret $10\times8 = 80$ bytes as $10$ doubles, there is a
commonly used serialization standard: protobuf (short for Protocol Buffers).

The data to be serialized is described in a schema file which is then passed to a
compiler (``protoc'', provided by the protobuf-toolkit) that generates ``message''
classes in the language of both server and client. By using these generated message
classes for serialization and deserialization, both the server and client can
exchange and reconstruct structured data.

This architecture provides a clean separation between simulation and optimization.
The simulation code remains entirely in \cpp\ and existing simulation code can be
used in a TCP-server with a few simple modifications while optimization and learning
logic can be developed in Python, making use of its many practical libraries. The
TCP interface is robust, flexible and requires minimal overhead.

\section{Results}
\label{section:results}

\begin{figure}[t]
	\fbox{
		\begin{minipage}{\textwidth - 4mm}
			\begin{minipage}{.5\textwidth}
				\subfloat[]{\raggedright{}\label{fig:comparison}}\vspace{-0.25cm}
				\begin{center}
					\includegraphics[width=\textwidth]
					{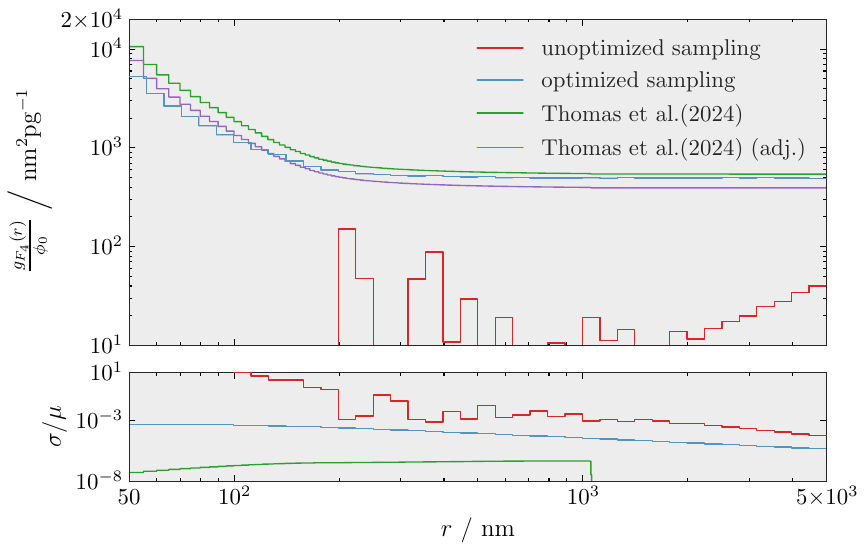}
				\end{center}
			\end{minipage}%
			\begin{minipage}{.5\textwidth}
				\subfloat[]{\raggedright{}\label{fig:q_comparison}}\vspace{-0.395cm}
				\begin{center}
					\includegraphics[width=0.9175\textwidth]
					{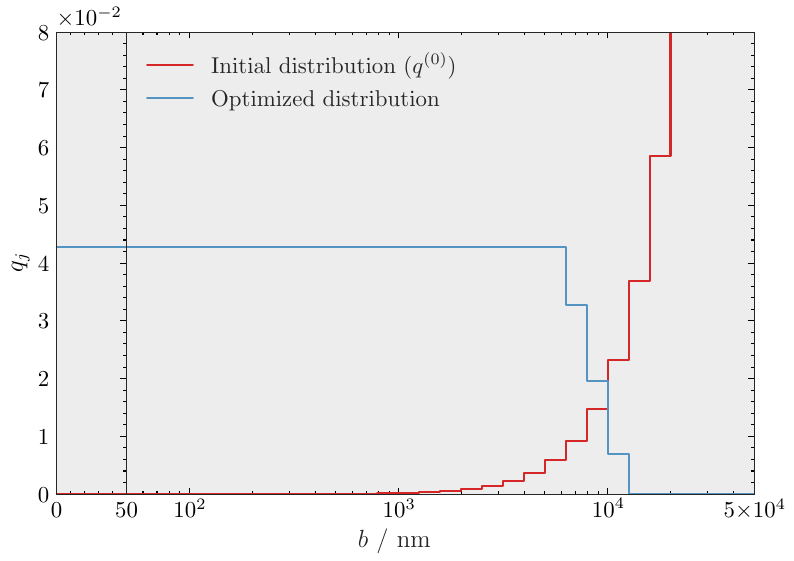}
				\end{center}
			\end{minipage}
			\caption{
				\textbf{\protect\subref{fig:comparison}}
				Comparison of the cluster dose as a function of radial distance from
				the NP (eq.~\ref{eq:exp-mean}) for the optimized importance function
				(blue curve), the ``analog'' computation of eq.~\ref{eq:exp-mean}
				(red curve) as well as data taken from \citeintext{Thomas_2024}
				in original (green curve) and fluence-adjusted (purple curve) form
				for comparison.
				The shaded area corresponds to a sampling uncertainty of one standard
				deviation $\sigma$. The bottom plot displays the corresponding
				relative uncertainties as $\sigma/\mu$.
				\textbf{\protect\subref{fig:q_comparison}}
				Comparison of the initial (red) and final (blue) distribution function that
				were used to generate the cluster doses in \protect\subref{fig:comparison}.
				The $y$-axis has been cut off for legibility omitting higher values of
				the initial importance distribution. The initial
				distribution is, however, identical so $p_j$ the numerical values of which
				can be viewed in table~\ref{tab:p_j}.
			}
			\label{fig:comparisons}
		\end{minipage}
	}
\end{figure}

Fig.~\subref*{fig:comparison} shows the shell-volume averaged cluster dose
$\mean{\gff}{r_i}$ normalized to the primary photon fluence as a function of radial
distance from the NP center with (blue) and without (red) the use of importance
sampling. The number of primary photons simulated is $10^9$ in either case.
Corresponding data from a two-step simulation \cite{Thomas_2024} is displayed for
reference (green). Due to its large simulation volume the setup used by these
authors allows for more secondary photons contributing and therefore an integral
fluence $1.38$ times the primary photon fluence due to photons scattering back into
the scoring volume was reported. For more accurate comparison, this reference data
has been adjusted for this factor (purple curve). The cluster dose at every
optimization iteration has been calculated with $N = 10^6$ (i.e. $\lfloor q_j
	N\rfloor$ primary particles for an annulus $j$) primary photons.

Without the use of any variance reduction technique the used number of primary
particles is far from sufficient; in fact the fluence incident on the inner shells
is so small that often no ionization clusters are expected to be scored at all (see
table~\ref{tab:p_j}). While the $p_j$ are orders of magnitude smaller towards inner
shells, their contribution is relevant to the computation of the shell-volume
averaged cluster dose $\mean{\gff}{r_i}$ (see eqs.~\ref{eq:exp-mean} and
\ref{eq:exp-mean-imp}). Given this suppression, the number of $10^9$ primary
particles is often insufficient to produce any interaction at all. This leads to
lower values of cluster dose even at larger radii.

The gold NP\textquotesingle s influence on ionization cluster generation is largely
confined to distances of up to $200\,\nm$ from the NP surface. The cluster dose in
shells beyond that range of immediate influence of the NP is largely made up by the
background contribution from photons interacting in water \cite{Thomas_2024}.

In the range of immediate NP influence the cluster dose obtained with the importance
sampling approach using the optimized importance function is in good agreement with
the fluence adjusted reference data. Beyond that range, up to a distance from the NP
center of $1\,\um$ the cluster dose obtained from importance sampling surpasses the
adjusted reference data. Notably, it shows agreement with the unmodified reference
data, which is larger by the fluence factor of $1.38$. Since, however, no physical
explanation of this behavior is apparent, the observed agreement with the unmodified
reference data is regarded as coincidental, and the increase is attributed to other,
as yet unidentified, factors.

\begin{figure}[t]
	\centering
	\fbox{
		\begin{minipage}{0.667\textwidth}
			\vspace{5 mm}
			\begin{center}
				\includegraphics[width=0.75
					\textwidth]{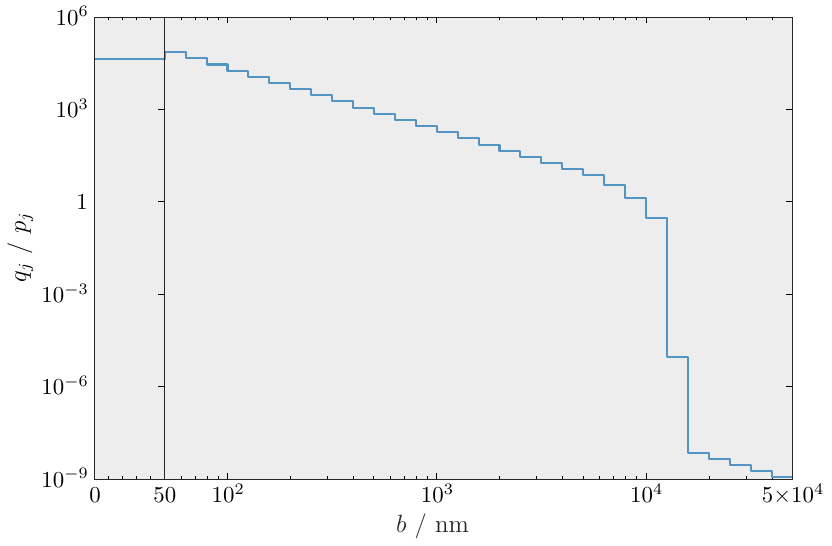}
			\end{center}
			\caption{
			Final importance function divided by initial importance distribution
			($q_J^{(0)} = p_j$). This is equal to the inverse likelihood ratio
			and also $\propto q_j / A_j$, the probability mass per area of the
			corresponding annulus.
			}
			\label{fig:q_over_p}
		\end{minipage}
	}
\end{figure}

Fig.~\subref*{fig:q_comparison} displays the initial as well as the optimized
importance distribution. The latter shifts weights towards the central axis (low
impact parameters). Up to impact parameters of $500\,\nm$, the probability masses
are equal; given the uneven bin-spacing, this corresponds to differing spatial
densities of primary particle position generation. Fig.~\ref{fig:q_over_p} shows the
quotient $q_j / p_j$ to show the relative change in probability masses. This
quantity is proportional to $q_j / A_j$---the probability mass of the importance
function per annulus area for an annulus $j$ (see eq.~\ref{eq-a:masses} in the
appendix)---and hence a measure proportional to the number of primary particles per
area.

For the first annulus (matched to the NP dimensions) the number of primary particles
generated by sampling from the optimized importance function exceeds the number of
primary particles generated using the unoptimized sampling by a factor of roughly
$4.3\smalltimes 10^{4}$ and for the second annulus, this factor reaches roughly $7.3
	\smalltimes 10^{4}$. The subsequent fall off roughly follows $1/b$ (this is
dominated by the annulus-area proportionality of $p$) and at impact parameters of
$\gtrsim 10\,\um$ the importance distribution $q$ suddenly decreases by some orders
of magnitude.

\begin{figure}[ht]
	\fbox{
		\begin{minipage}{\textwidth - 4mm}
			\begin{minipage}{.5\textwidth}
				\subfloat[]{\raggedright{}\label{fig:comparison-cumulative}}\vspace{-0.14cm}
				\begin{center}
					\includegraphics[width=0.895\textwidth]{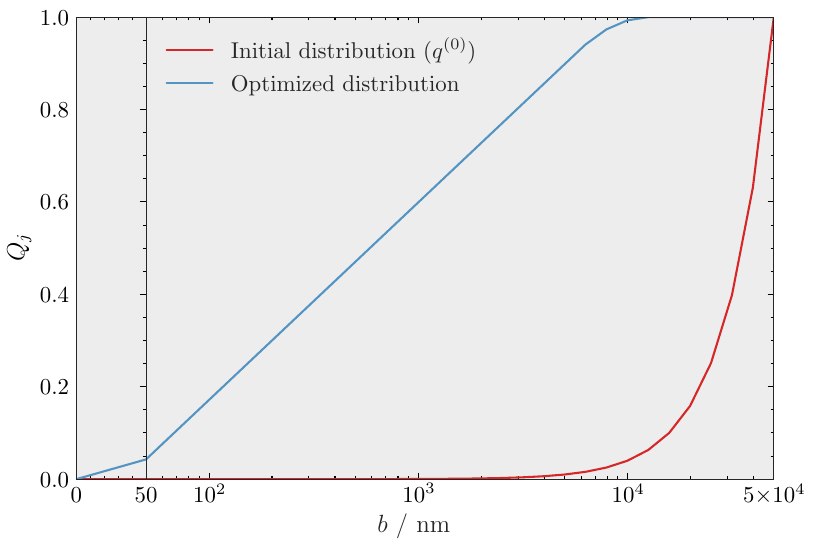}
				\end{center}
			\end{minipage}%
			\begin{minipage}{.5\textwidth}
				\subfloat[]{\raggedright{}\label{fig:comparison-heatmap}}\vspace{-0.35cm}
				\begin{center}
					\includegraphics[width=\textwidth]{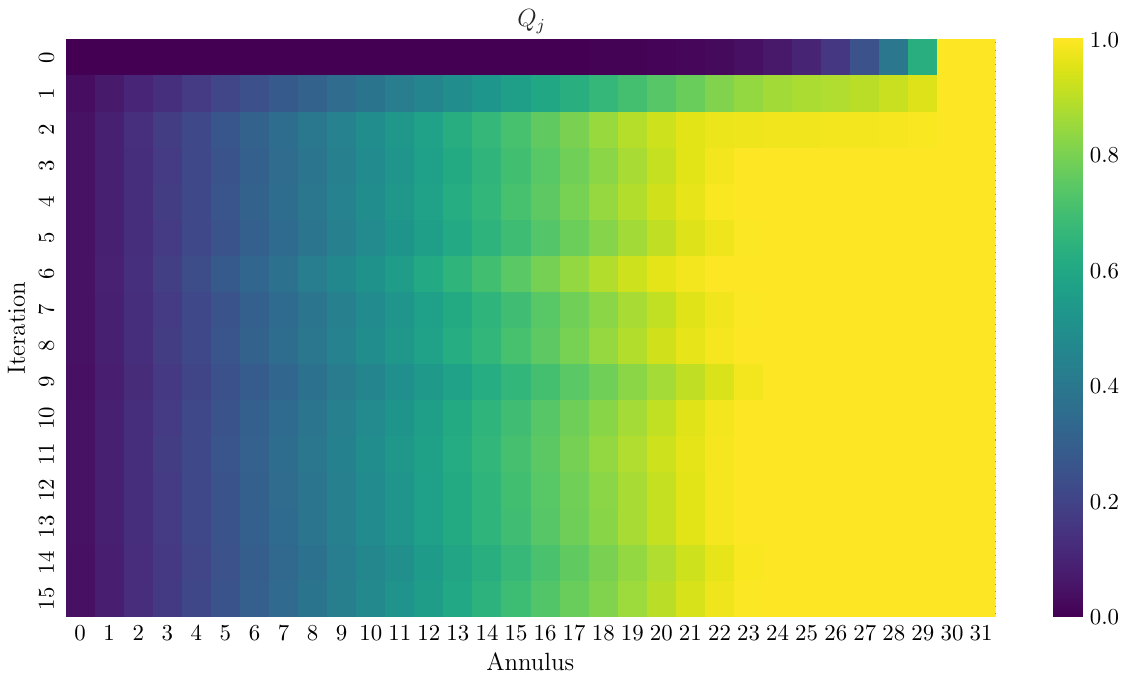}
				\end{center}
			\end{minipage}
			\caption{
			\textbf{\protect\subref{fig:comparison-cumulative}} Cumulative probability
			functions (CDFs) corresponding the data in fig.~\protect\subref*{fig:q_comparison}
			for the initial (red) and final (blue) optimization step.
			\textbf{\protect\subref{fig:comparison-heatmap}} Heatmap-display of
			the CDFs for all iterations beginning with the initial CDF (top row)
			to the last iteration (bottom row).
			Note that 1. the values are annulus-wise descrete for better legibility and 
            2. the bin widths are not
			to scale; while the logarithmic bins edges are equidistant on a
			logarithmic scale, the first annulus covers impact parameters between
			$[0, 50\,\nm)$.
			}
			\label{fig:comparisons-cumulative}
		\end{minipage}
	}
\end{figure}

In addition fig.~\ref{fig:comparisons-cumulative} shows the cumulative probability
functions (CDFs, see eq.~\ref{eq-b:CDF}) corresponding to the probability
distributions in figs.~\subref*{fig:q_comparison}
(fig.~\subref*{fig:comparison-cumulative}). Fig.~\subref*{fig:comparison-heatmap}
displays the initial CDF $Q^{(0)}$ and the CDFs after each iteration ($Q^{(k)}$,
with $k=1,\dots,20$, top to bottom) in a heatmap. This heatmap serves as a visual
representation of the convergence of the iterative process: Already after 2-3
iterations the importance function fluctuates around a distribution that is
considered stationary. Slight outliers (such as $Q^{(6)}$ or $Q^{(9)}$) occur, the
algorithm subsequently, returns to this stationary distribution.

\begin{figure}[t]
	\centering
	\fbox{
		\begin{minipage}{0.667\textwidth}
			\vspace{5 mm}
			\begin{center}
				\includegraphics[width=0.75
					\textwidth]{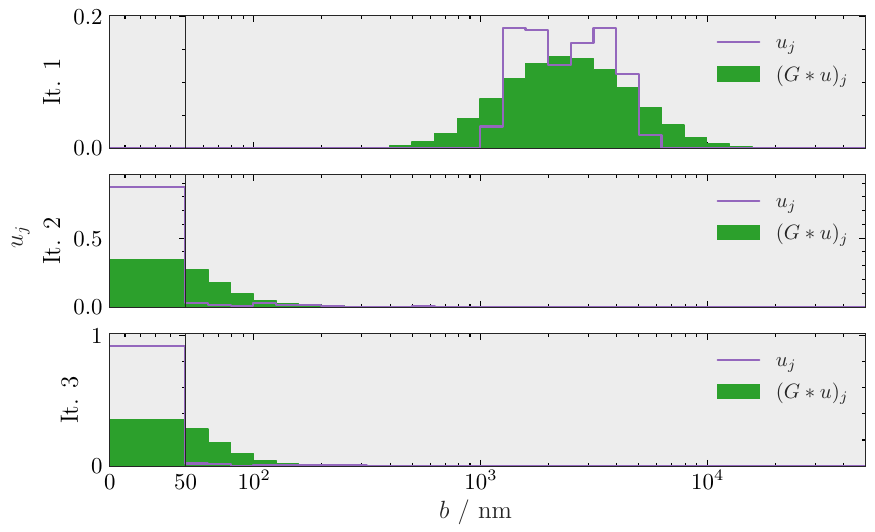}
			\end{center}
			\caption{
				Importance scores calculated from the data obtained in the first
				three iterations as well as the corresponding transformed scores
				(eq.~\ref{eq:conv}, parameter \mbox{$\sigma = 2$}).
			}
			\label{fig:u_j}
		\end{minipage}
	}
\end{figure}

The importance scores obtained from the first three iterations are displayed in
fig.~\ref{fig:u_j}. At this point they have converged almost entirely, with the
biggest importance attributed to the zeroth annulus which is matched to the NP
radius. This is consistent with the fact that photons impinging on the NP are more
likely to interact and produce secondary electrons.

\begin{figure}[t!]
	\fbox{
		\begin{minipage}{\textwidth - 4mm}
			\begin{minipage}{.5\textwidth}
				\subfloat[]{\raggedright{}\label{fig:metrics_efficiency}}\vspace{-0.35cm}
				\begin{center}
					\includegraphics[width=\textwidth]{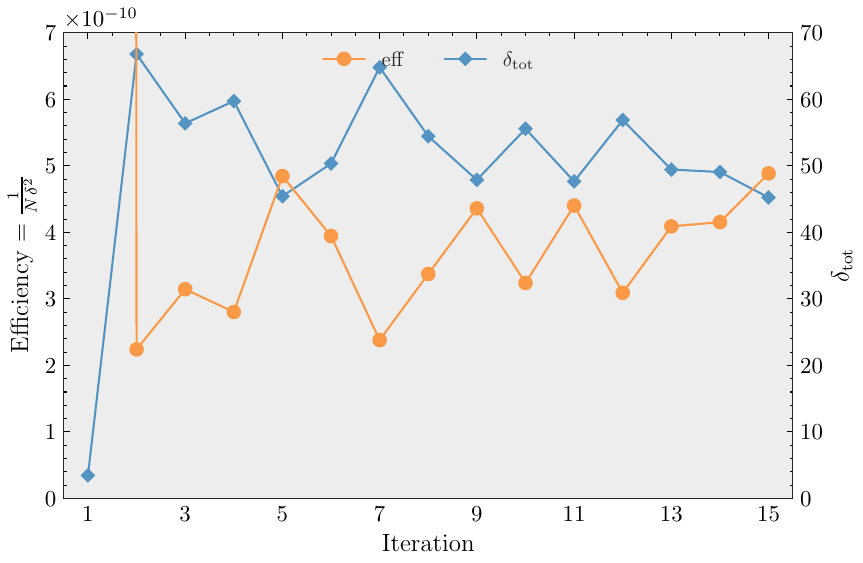}
				\end{center}
			\end{minipage}%
			\begin{minipage}{.5\textwidth}
				\subfloat[]{\raggedright{}\label{fig:metrics_W}}\vspace{-0.20cm}
				\begin{center}
					\includegraphics[width=0.97\textwidth]{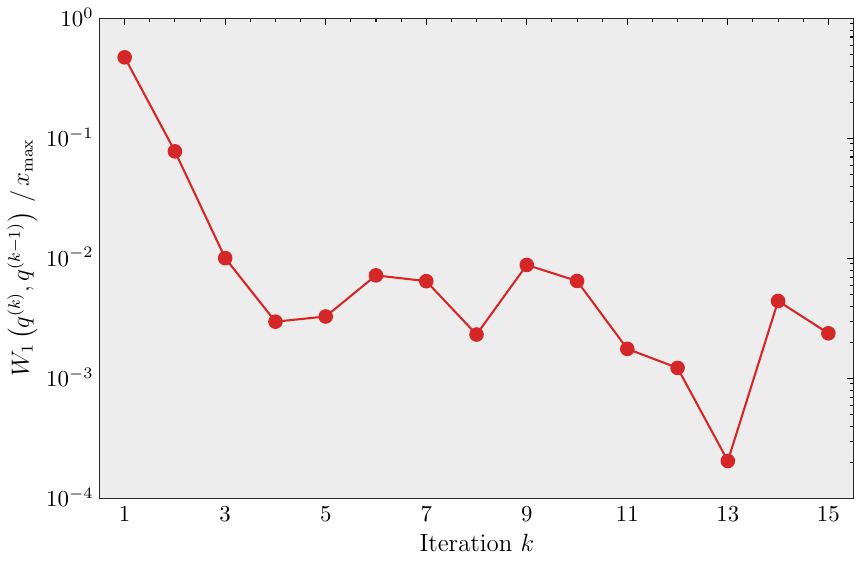}
				\end{center}
			\end{minipage}
			\vspace{-0.7mm}
			\begin{center}
				\begin{minipage}{.5\textwidth}
					\subfloat[]{\raggedright{}\label{fig:metrics_gt}}\vspace{-0.35cm}
					\begin{center}
						\includegraphics[width=\textwidth]{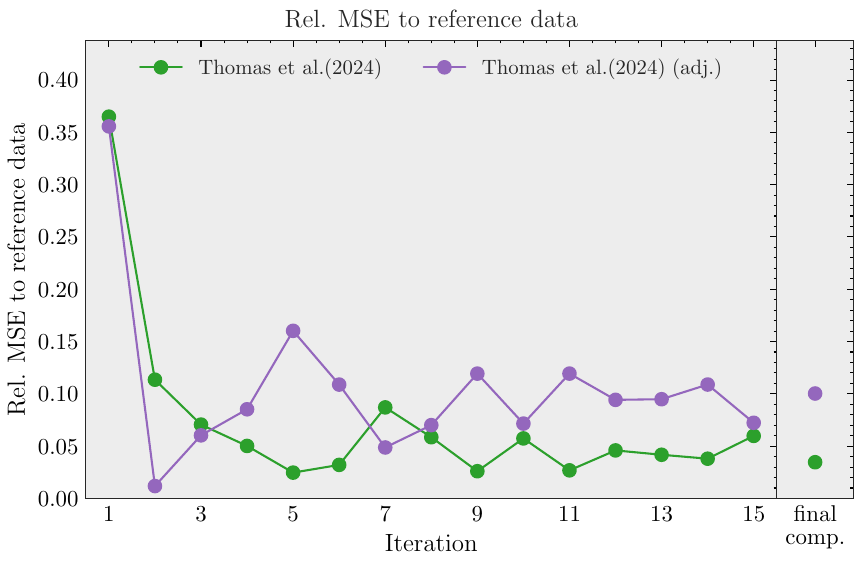}
					\end{center}
				\end{minipage}
			\end{center}
			\caption{
				Metrics used to evaluate optimization performance.
				\textbf{\protect\subref{fig:metrics_efficiency}} depicts the
				efficiency, expressed as $1/(N\delta^2)$ (left $y$-axis, the
				efficiency for the first iteration is $8.35\smalltimes10^{-8}$) and
				is cut for legibility) as well as
				the relative uncertainty $\delta$ in the entire scoring volume (right
				$y$-axis).
				\textbf{\protect\subref{fig:metrics_W}} measures the difference
				between an importance distribution and the importance distribution of
				the preceding iteration.
				\textbf{\protect\subref{fig:metrics_gt}}
				Mean squared error
				between the cluster dose in all scoring shells for the data simulated
				in an iteration and the cluster dose in the matching scoring volume
				taken from \citeintext{Thomas_2024}.
			}
			\label{fig:metrics}
		\end{minipage}
	}
\end{figure}

Fig.~\ref{fig:metrics} displays different metrics used to evaluate or monitor the
optimization procedure. Of special interest is the efficiency of the simulation
which can be quantified as $1/(N\delta^2)$, where $\delta$ is the scored standard
deviation relative to the mean; fig.~\subref*{fig:metrics_efficiency} displays how
the efficiency evolves during optimization. The efficiency is the largest after the
first iteration, this is, however, misleading as it is the result of zero-valued
cluster doses and zero-valued uncertainties \footnote{since there is no sampling
	uncertainty associated to no samples scored, the associated uncertainty is undefined
	and defaults to zero in this calculation.}. It increases and stabilizes around this
initial efficiency increase (approximately at $4\smalltimes10^{-10}$).

The scored uncertainty is made up of two components: an uncertainty resulting from a
limited number of primary particles contributing to the tally (epistemic
uncertainty) as well as the uncertainty of $\gff$ itself (aleatoric uncertainty).
The former component presumably dominates the cluster dose scored in the first
iteration: only very few particles contribute to scoring. This component is of
interest to the optimization carried out. After the first few iterations the latter
contribution pertaining to the stochasticity of the cluster dose itself possibly
overlays any further reduction in sampling uncertainty. This matter as well as the
appropriate optimization criterion are further addressed in the discussion section.

An alternative optimization metric is a measure for the change made to the
importance distribution after each iteration, namely the $W_1$ distance between an
importance distribution $q^{(k)}$ to the distribution from the prior iteration
$q^{(k-1)}$ (fig.~\subref*{fig:metrics_W}). It shows a sharp drop spanning the first
three iterations and stabilizes after an outlier at iteration 5. This can also be
seen in the heatmap plot of the CDFs (fig~\subref*{fig:comparison-heatmap}). After
this outlier and smaller following outliers, the algorithm fluctuates back to a
distribution resembling the final distribution.

Fig.~\subref*{fig:metrics_gt} shows the mean squared error (MSE) between the
adjusted reference data and the scored cluster dose relative to adjusted (purple)
and unadjusted (green) reference data. This information is not used as an
optimization metric or at all during optimization, but for validation only. For the
final comparison simulation, the relative MSE to the reference data is
$10\,\%$ and $3.5\,\%$ for the adjusted and unadjusted reference data,
respectively.

\section{Discussion}
\label{section:discussion}

\subsection{Main findings}
\label{subsection:findings}

The optimization method developed here addresses the complex problem of inferring an
optimal importance distribution for importance sampling in an active-learning, i.e.
data-driven approach. In the use case of the present study it converges quickly and
yields physically consistent results. Although the initialization of the importance
function as $q_j^{(0)} = p_j$ is far off from the final result, the method succeeds
in exploring undersampled regions and robustly identifies annuli matching the NP as
well as the scoring volume as main contributors.

Regarding the final importance distribution: as discussed, the definition of the
term ``optimal importance distribution'' is inherently debatable. What an optimal
importance distribution is, is encoded in the loss function
(eq.~\ref{eq:loss-final}) which is based on two heuristics: First, the fraction of
samples drawn from an annulus should be proportional to the contribution that
primary photons originating from that annulus yield (this is similar to the
heuristic from \citeintext{RobertCasella_2004}). Second, the fluctuation between
adjacent bins should be limited in regions that are identical in material, which is
implemented using the regularization term. Any importance distribution that
minimizes such a loss function is ``optimal'' in that sense and any importance
distribution is bound to yield correct results, provided the simulation is run long
enough and importance weights do not vanish in regions that are relevant to the
result.

A more relevant measure of success of the present method is how it compares to other
methods of variance reduction, such as the two-step method that has been used to
generate the reference data. The reference data is considered more accurate because
it more plausibly reproduces the expected physical behavior, at larger distances
from the NP, in particular the constant background contribution. In addition, its
statistical uncertainty is lower by up to five orders of magnitude. These advantages
come at a cost, however: generating the data required several days to weeks of
computation rather than hours and involved multiple separate simulations
(background, NP, and a ``water'' NP). Although improvements remain necessary, the
method presented here can already be regarded as a successful proof-of-principle.

The optimized importance function $q$ (fig.~\subref*{fig:q_comparison}) is used for
a simulation with $10^9$ primary photons and compared to the results obtained using
the physical (or unoptimized) distribution $p$. The resulting
(fig.~\subref*{fig:comparison}) shell-volume averaged cluster doses
$\mean{\gff}{r_i}$ normalized to the primary fluence as a function of radial
distance from the NP center are compared to previously published reference data.

For the number of primary particles simulated the use of the unoptimized
distribution does not result in sufficient fluence near the gold NP, as elaborated
above (see table~\ref{tab:p_j}). The use of the optimized importance distribution
increases the fluence in the vicinity of the NP by over four orders of magnitude
(this increase is proportional to $q_j/p_j$, depicted in fig.~\ref{fig:q_over_p})
and yields results closer to the reference data.

The overall smaller simulation volume does not allow for the inclusion of all
photons that might scatter back into the scoring volume. Their contribution amounts
to an increase of the primary photon fluence of $38\,\%$ \cite{Thomas_2024} and the
reference data has been adjusted to allow for direct comparison. While the data
obtained using the optimized importance distribution aligns well with the adjusted
reference data in the influence region of the gold NP ($r\lesssim 200\,\nm$), it
overestimates the cluster dose in the region that is dominated by the background
contribution ($r\gtrsim 200\,\nm$), as discussed in section~\ref{section:results}.

The mean squared error between the obtained cluster dose in the entire scoring
volume and the cluster dose from adjusted and unadjusted reference data in the same
volume is $10\,\%$ and $3.5\,\%$ for the adjusted and unadjusted, respectively,
see fig.~\subref*{fig:metrics_gt}. Major contribution is deemed to be the
overestimation of the background, in fact, the relative MSE to the reference data
	is smaller for the unadjusted reference data.


A central challenge in developing the method was the definition of appropriate
metrics for evaluating convergence of the optimization. Two different approaches
were explored. The first relied on an uncertainty-based measure, where the
efficiency was quantified as $1/(N\delta^2)$, with $\delta$ denoting the scored
relative standard deviation, see fig.~\subref*{fig:metrics_efficiency}.
Following a large initial value at the first iteration that can be seen as an
artifact attributed to the way the algorithm handles no scored cluster dose, the
efficiency increases to values of approximately $4\smalltimes10^{-10}$).

The fluctuation in the relative uncertainty $\delta$ is---in part---a result of the
number of primary photons used for optimization ($N=10^6$). While the optimization
can be done with more extensive simulations, the number of primary particles is
restricted by design: the purpose of the simulations during optimization is merely
to obtain an optimal importance function for a ``proper'' simulation (here the final
comparison). Increasing the number of primary particles during optimization beyond
what is necessary would defeat the purpose of variance reduction.

As outlined in the results section the uncertainty has two components: an aleatoric
component, related to the intrinsic fluctuations of $\gff$ and an epistemic
component, related to the sampling process. It is this latter component that is of
relevance to the optimization process and in principle, this component may be
isolated using the law of total variance. This would, however, require reliable
estimates for the mean and variance of the scored cluster dose values given a
certain impact parameter which would come at a computational cost
that---again---ultimately defies the purpose of this method as a variance reduction
scheme.

As a second approach, convergence was evaluated in terms of the stability of the
obtained distributions themselves. For this, the $W_1$ distance was calculated
between the importance distribution of successive iterations, see
fig.~\subref*{fig:metrics_W}. This metric provides a direct measure of how much the
importance distributions change from one step to the next. The results reveal a
clear trend towards convergence: while some fluctuations remain, the difference
between successive distributions are minor. Fig.~\subref*{fig:comparison-heatmap}
shows that the corresponding distributions vary around a stationary distribution.



\subsection{Relevance}
\label{subsection:relevance}

The optimization strategy developed in this work allows for application of
importance sampling in situations where it is not straightforward to define an
appropriate importance function. In this work, the method has been applied to the
calculation of the radial dependence of the $F_4$-cluster dose around a gold NP. It
makes use of the inherent symmetries of the system and is in its present form
directly applicable to other such systems. Nevertheless, the underlying concept can
be extended straightforwardly to more general geometries.

In the broader field of radiation dosimetry, methods that fall under the umbrella of
artificial intelligence are increasingly employed \cite{Hu_2023, Irannejad_2024,
	Hou_2025, Schwarze_2025}. To the best of the authors\textquotesingle \ knowledge,
however, such approaches have not yet found use in nanodosimetry.

This may be explained, at least in part, by the characteristics of the respective
domains: machine learning techniques have demonstrated their greatest success in
high-dimensional tasks such as the prediction of dose maps. Nanodosimetry, on the
other hand, typically involves comparatively low-dimensional data. Moreover,
track-structure computations in general and nanodosimetric calculations in
particular typically involve tasks that demand high accuracy and predictions based
purely on machine learning models may not always be sufficiently reliable for this
purpose.

The present method avoids these limitations: while the optimization is
heuristic-driven, these heuristics serve solely to construct a better-performing
importance function, effectively accelerating convergence. The data are generated
using simulation.

\subsection{Caveats and final remarks}
\label{subsection:caveats}

The choice of the loss function $\mathcal{L}$ implicitly defines what constitutes an
optimal importance distribution. While this flexibility allows the method to
accommodate specific objectives that may be relevant for different use cases, it
requires ensuring that the formulation of $\mathcal{L}$ reflects meaningful
criteria.

An optimal importance function also depends on the quantity of interest. Since the
generation of an $F_4$-cluster requires four or more ionizations, by definition,
their occurrence is more localized than the conventional dose, which is the
mean of \textit{all} energies imparted to a certain mass. Accordingly, the
functional relationship between the impact parameter and the dose might look
different and the optimization procedure might yield slightly different results.

The impact parameter $b$ as well as the distance from the origin $r$ to score
cluster dose cover fairly large scales. To keep the number of annuli and shells
under control, their spacing has been chosen to be logarithmic. While this allows
for decent resolution closer to the NP, annuli/scoring volumes increase
exponentially. This allows for the possibility of feature loss and the spacing
chosen here is a trade-off between large coverage and high resolution.

Regarding $\gff$ purely as a function of radial distance $r$ from the NP center
disregards anisotropies of cluster production. For the absorbed dose, the impact of
such anisotropies has been determined to be in the few-percent range for CPE
conditions \cite{Derrien_2023, Rabus_2024b}.

\section{Conclusions and Outlook}
\label{section:conclusions}

This work has described a novel method of using importance sampling for Monte Carlo
estimators based on obtained data samples rather than geometric assumptions. As
such, it may be applied to different scenarios that require more efficient use of
sample generation.

The method presented here is a first demonstration. Its main benefit is that it
formally does not introduce a bias to the estimator---this is as long as the
optimization does not lead to vanishing probabilities in regions with relevant
contributions. Once an optimal importance distribution is found, it can be reused.

Multi-step paradigms, on the other hand, usually require making some assumptions
such as invariance of the radiation field under rescaling (for methods shrinking
phase space files) or the absence of synergistic effects of secondary particles
originating from the same primary particle. A strength of these paradigms is their
ability to capture more photons that are scattered back into the simulation volume
as the volume of the first simulation can be chosen almost arbitrarily wide. On the
other hand, the (repeated) use of phase space files may reproduce statistical biases
from the recorded phase space file such as the fixed association between position
and energy.

To the authors\textquotesingle \ knowledge no prior works made use of interactive
interfacing of MC code with python scripts. This method of interfacing may well be
of use for other applications, especially those using machine learning.

Future work will need to address the issue of a proper convergence criterion. Since
the point of importance sampling is to increase the efficiency of sampling, a
sampling variance-based measure appears to be the obvious choice. This requires the
ability to decompose the obtained sampling uncertainty into the part that stems from
(insufficient) sampling and the part that reflects the stochasticity of the
observable itself. Prerequisite is the accurate estimation of the latter component.


\section*{Acknowledgements}
\label{section:acknowledgements}

The authors express their gratitude to the dedicated team of the High Performance
Cluster of the German National Metrology Institute (PTB) for their ongoing support
throughout the production of the data.

This project is part of the programme ``Metrology for Artificial Intelligence in
Medicine'' (M4AIM), which is funded by the Federal Ministry for Economic Affairs and
Energy (BMWE) within the scope of the ``QI-Digital'' initiative.

\footnotesize
\bibliographystyle{apalike}
\bibliography{references}




\begin{appendix}
\normalsize
\renewcommand{\thesection}{\Alph{section}}
\titleformat{\section}{\normalfont\Large\bfseries}{Appendix \thesection}{1em}{}
\renewcommand{\theequation}{\Alph{section}.\arabic{equation}}
\numberwithin{equation}{section}


\section[Appendix \thesection]{}
\label{sec:appendixA}

\subsection{Annulus- and shell-averaged cluster dose}
\label{subsection:averaging}

The $F_4$-cluster dose $\gff$ at a point $\rr$ can be viewed as the expectation of
the---not analytically accessible---function $\gff$ over all possible starting
positions of the primary photons $\xx\sim p^{\scriptstyle\mathrm{phys.}}$:
\begin{align}
	\frac{\gff(\rr)}{\phi_0}\ =\ \frac{1}{\phi_0}\underset{\xx\sim p^{\scriptstyle\mathrm{phys.}}}{\mathbb{E}}
	\Big[\gff(\rr|\xx)\Big]\ =\ \frac{1}{\phi_0}\int\dd A\
	p^{\scriptstyle\mathrm{phys.}}(\xx)\gff(\rr|\xx). \label{eq-a:exp}
\end{align}

Here, $p^{\scriptstyle\mathrm{phys.}}$ is the uniform probability density function
(PDF) representative of the physical distribution of the starting positions with
coordinates $\xx$. The source is modelled so that primary particles are uniformly
generated on a disk of area $A$ that is in the $x$-$y$-plane and located at $z =
	-d_\mathrm{src}$ (with $d_\mathrm{src} = 100\,\um$), see region `$\mathbf{B}$'
fig.~\ref{fig:setup} as well as fig.\ref{fig:impact} in the main text.

The central axis from the center of the source $(0,0,-d_\mathrm{src})$ to the origin
of the coordinate system (the center of the NP) is a symmetry axis and since the
source is confined to a plane, the starting position can effectively be
characterized by the distance to the origin of the source, which will be referred to
as
\begin{align}
	b \equiv \sqrt{x^2 + x^2}.
\end{align}

Presuming the absence of anisotropies in the ionization cluster dose, $\gff$ can be
considered as a function of the distance $r \equiv |\rr|$ from the NP center only.

The PDF $p^{\scriptstyle\mathrm{phys.}}$ over the impact parameter in
eq.~\ref{eq-a:exp} is
\begin{align}
	p^{\scriptstyle\mathrm{phys.}}(\xx) = \frac{\delta(z +
		d_\mathrm{src})}{A}\quad\mathrm{for}\ b\ \in\ [0, b_\mathrm{max})
	\qquad (\mathrm{and}\ 0\ \mathrm{else}) \label{eq-a:pdf-3d}
\end{align}
where $b_\mathrm{max}$ is the maximum impact parameter for the source and $\delta$
is the Dirac-$\delta$ function. Using the symmetry around the $z$-axis
eq.~\ref{eq-a:exp} can be transformed in cylindrical coordinates:
\begin{align}
	\frac{\gff(r)}{\phi_0}
	\ =\ \frac{1}{\phi_0}\int\dd\xx\ p^{\scriptstyle\mathrm{phys.}}(\xx)\gff(r|b)
	\ =\ \frac{2\pi}{A\phi_0}\int\dd b\ b\,\gff(r|b). \label{eq-a:exp-cyl}
\end{align}

To facilitate optimization this work approximates the PDF with the superscript
`$\mathrm{phys.}$' as \textit{piece-wise constant} over $n_b$ single annuli $[b_j,
	b_{j+1})$, where $b_0 = 0$ and $b_{n_b} = b_\mathrm{max}$. For any given annulus,
primary particles are uniformly generated within its confines. This stratified
sampling allows for more efficient exploration of the impact parameter space, as it
effectively reduces the continuous distribution to a discrete set of
\textit{probability masses}: One can average the probability density function over
the considered bins
\begin{align}
	p_j \equiv \int_{b_j}^{b_{j+1}}\bdd b\, 2\pi\, b\, p^{\scriptstyle\mathrm{phys.}}(\xx)
	= \frac{\pi(b_{j+1}^2 - b_j^2)}{A}
	= \frac{A_j}{A}, \label{eq-a:masses}
\end{align}
where the $p_j$ are their associated probability masses. Evidently the $p_j$ sum up
to $1$.
The integral in eq.~\ref{eq-a:exp-cyl} approximated with the piece-wise PDF is
\begin{align}
	\frac{\gff(r)}{\phi_0}
	\                                             & =\ \frac{2\pi}{A\phi_0}\int\dd b\ b\,\gff(r|b)                                    \\
	\intertext{and can then be solved annulus-wise:}
	                                              & =\ \frac{2\pi}{A\phi_0}\sum_{j=0}^{n_b-1}\int_{b_j}^{b_{j+1}}\bdd b\ b\,\gff(r|b) \\
	                                              & =\ \frac{1}{\phi_0}\sum_{j=0}^{n_b-1}\frac{A_j}{A}\mean{\gff(r)}{b_j}             \\
	(\ref{eq-a:masses})\quad \Longrightarrow\quad & =\ \frac{1}{\phi_0}\sum_{j=0}^{n_b-1} p_j\,\mean{\gff(r)}{b_j}
	\label{eq-a:exp-annulus}
\end{align}
where $\mean{\gff(r)}{b_j}$ has been implicitly defined as the mean cluster dose
in the annulus defined by $\mathcal{A}_j := \{(r,\phi)\in [b_j,
	b_{j+1})\times[0,2\pi)\}$ with an area $A_j \equiv |\mathcal{A}_j|$\footnote{The
	term ``annulus mean'' is chosen over the term ``annulus average'' in order to avoid
	the possible confusion with ``an average over annuli''.}. For any quantity $Q$
that depends on $b$, the mean is:
\begin{align}
	\mean{Q}{b_j}
	\equiv \frac{1}{A_j} \int_{\mathcal{A}_j}\dd A\,Q
	= \frac{2\pi}{A_j} \int_{b_j}^{b_{j+1}} \bdd b\, b\,Q. \label{eq-a:mean_b}
\end{align}

The results obtained from simulation (outlined in
section~\ref{subsection:simulation}) is the mean cluster dose in a shell defined by
$\mathcal{V}_i := \{(r,\theta,\phi)\in [r_i, r_{i+1})\times[0,\pi)\times[0,2\pi)\}$
with volume $V_i \equiv |\mathcal{V}_i|$. A volume-mean is then:
\begin{align}
	\mean{Q}{r_i} \equiv \frac{1}{V_i} \int_{\mathcal{V}_j}\dd V\,Q
	 & = \frac{4\pi}{V_i} \int_{r_i}^{r_{i+1}} \bdd r\, r^2\,Q , \label{eq-a:mean_r}
\end{align}
Applying the volume-average to eq.~\ref{eq-a:exp-annulus} one finds:
\begin{align}
	\boxed{
		\frac{\mean{\gff}{r_i}}{\phi_0}
		\ =\ \frac{1}{\phi_0}\sum_{j=0}^{n_b-1} p_j\,\mean{\gff}{r_i, b_j}
		\label{eq-a:exp-mean}
	}
\end{align}





\textbf{Note:}
In radiation physics it is more prevalent to consider \textit{differential
	quantities} rather than mean quantities such as the ones in
eqs.~\ref{eq-a:mean_b}~and~\ref{eq-a:mean_r}. The---arguably less elegant---formalism
used here is, however, consistent with the use of expectation values, as is done
for importance sampling, since here $p(b)$ can be interpreted as a probability
density.

\subsection{Importance sampling}
\label{subsection:importance}

Eq.~\ref{eq-a:exp-cyl} can be solved using importance sampling using an importance
function $q(\xx)$, a PDF that has shares the same support of
$p^{\scriptstyle\mathrm{phys.}}(\xx)$. The use of averaged quantities, however, it
is sufficient to think of the importance function as piece-wise constant only. And
it is hence fully characterized by its probability masses, equivalent to those
defined in eq.~\ref{eq-a:masses}:
\begin{align}
	q_j \equiv \int_{b_j}^{b_{j+1}}\bdd b\, 2\pi\, b\, q(\xx). \label{eq-a:masses-q}
\end{align}

\begin{align}
	\frac{\gff(r)}{\phi_0}
	\  & =\ \frac{2\pi}{A\phi_0}\int\dd b\ b\,\frac{q(\xx)}{q(\xx)}\gff(r|b)                 \\
	\intertext{Averaged over annuli this yields}
	   & =\ \frac{2\pi}{A\phi_0}\sum_{j=0}^{n_b-1}\frac{q_j}{q_j}\int_{b_j}^{b_{j+1}}\bdd b\
	b\,\gff(r|b)                                                                             \\
	   & =\ \frac{1}{\phi_0}\sum_{j=0}^{n_b-1}q_j\frac{p_j}{q_j}\mean{\gff(r)}{b_j}
\end{align}
Averaged over shell volumes this yields:
\begin{align}
	\boxed{
		\frac{\mean{\gff}{r_i}}{\phi_0}
		\ =\ \frac{1}{\phi_0}\sum_{j=0}^{n_b-1}q_j\frac{p_j}{q_j} \mean{\gff}{r_i, b_j}
		\label{eq-a:exp-mean-imp}
	}
\end{align}

Note that eq.~\ref{eq-a:exp-mean-imp} and eq.~\ref{eq-a:exp-mean} are identical. The
$q_j$ cancel out and $\mean{\gff}{r_i, b_j}$ as a mean does not depend on the number
of primaries used for optimization.






\section[Appendix \thesection]{}
\label{sec:appendixB}

\subsection{Wasserstein-1 distance}
\label{subsection:W1}

The Wasserstein-1 distance $W_1$ between two discrete probability distributions
$q_1$ and $q_2$ can be calculated from the corresponding cumulative distribution
functions (CDFs). For the piecewise linear distributions these are
\begin{align}
    Q(b) = \underbrace{\sum_{k=0}^{j-1} q_k}_{\equiv Q_{j-1}} + (b-b_j)\frac{q^j}{\Delta b_j}
    \quad \mathrm{if}\quad & b\in[b_j,b_{j+1}) \label{eq-b:CDF}
\end{align}
where $\Delta b_j \equiv b_{j+1} - b_j$. The $W_1$ distance can then be obtained via
\begin{align*}
	W_1(q_1, q_2) & = \int\dd b\, \Big|Q_1(b) - Q_2(b)\Big|
	= \sum_{j=0}^{n_b-1}\int_{b_j}^{b_{j+1}}\bdd b\,
	\Big|Q_1^j - Q_2^j+ (b-b_j)\frac{q_1^j - q_2^j}{\Delta b_j} \Big| =
	\sum_{j=0}^{n_b-1} I_j.
\end{align*}
The integral $I_j$ can be solved with the substitution
\begin{align*}
	u & \equiv Q_1^j - Q_2^j+ (b-b_j)\frac{q_1^j - q_2^j}{\Delta b_j} \quad\Longrightarrow\quad \dd u = \frac{q_1^j - q_2^j}{\Delta b_j}\dd b
\end{align*}

\begin{align*}
	I_j & = \frac{\Delta b_j}{q_1^j - q_2^j} \int_{u(b_j)}^{u(b_{j+1})}\bdd u\, |u| = \frac{\Delta b_j}{q_1^j - q_2^j} \left.\frac{u|u|}{2}\right|_{u(b_j)}^{u(b_{j+1})}
\end{align*}
with
\begin{align*}
	u(b_j)                       & = Q_1^j - Q_2^j                   \\
	\mathrm{and}\quad u(b_{j+1}) & = Q_1^j - Q_2^j + (q_1^j - q_2^j)
\end{align*}

\end{appendix}


\end{document}